\documentclass{article}
\usepackage{graphicx}
\begin{document}
\title{Periodic mass extinctions and the Planet X model reconsidered\\
\vspace{.5in}
Daniel P. Whitmire}

\author{Department of Mathematics, The University of Arkansas, Fayetteville, AR}

\maketitle

\begin{abstract}
The 27 Myr periodicity in the fossil extinction record has been confirmed in modern data bases dating back 500 Myr, which
is twice the time interval of the original
analysis from thirty years ago. The surprising regularity of this period has been used to reject the Nemesis model. A second model based on the sun's vertical galactic oscillations has been challenged on the basis of an inconsistency in period and phasing. The third astronomical model originally proposed to explain the periodicity is the Planet X model in which the period is associated with the perihelion precession of the inclined orbit of a trans-Neptunian planet. Recently, and unrelated to mass extinctions, a trans-Neptunian super-Earth planet has been proposed to explain the observation that the  inner Oort cloud objects Sedna and 2012VP113  have perihelia that lie near the ecliptic plane. In this Letter we reconsider the Planet X model  in light of the confluence of the modern palaeontological and outer solar system dynamical evidence.

\noindent \bf{Key Words}: astrobiology - planets and satellites - Kuiper belt: general - comets: general
\end{abstract}

\section*{Background}

The discovery of an apparent 26 Myr period in mass extinctions by Raup \& Sepkoski (1984) invited the proposal of three astronomical models. The
Nemesis model postulated that the extinction period was associated with comet showers that were triggered when a solar companion object in an eccentric orbit scattered comets when passing through it's perihelion point located in the inner Oort cloud. One version of the Nemesis model (Davis et al., 1984) assumed that the companion was a common M-dwarf star. This version was ruled out by full sky surveys, the latest of which is the NASA WISE mid-IR survey (Luhman, 2014). However, the other version of Nemesis (Whitmire and Jackson, 1984) assumed the companion was a sub-stellar object with a mass between 0.07 M$_{\odot}$ and $2\times 10^{-4}  $M$_{\odot}$. The lower end of this mass range is not ruled out by WISE or any other
full sky survey given that the object would likely be closer to aphelion today.

Using the revised 2012 geological timescale, Melott \& Bambach (2014, 2013, 2010) have given a period analysis based on two independent sets of data with increased taxonomic resolution. Their analysis confirmed a 27 Myr period over the entire Phanerozoic (500 Myr), which is twice the time interval
of the original Raup and Sepkoski analysis.  Using two different statistical tests they give period p-values of 0.004 and 0.02, which is more significant than  the original 0.05 given by Raup and Sepkoski.  As they note, there is not a local
 maxima in extinction at each 27 Myr interval and there are some peaks that lie between intervals. This is expected if
comet showers are the proximate
 cause of the peaks. We note that large KT-size comet impacts are not necessarily the only or even the primary mechanism for mass extinctions.
  Dust and numerous smaller impacts from the breakup of giant comets is a more likely scenario (Napier, 2015).
 
 Using the new geological timescale the spectrum of interval lengths does not show a 27 Myr period,
eliminating this as a possible
 artifact producing the original periodicity (Melott and Bambach, 2014).
 The period appears to be not only real but also highly regular with a variation of no  more that  2 - 10\% over 500 Myr.
If correct, as they
 point out, this rules out both versions of the Nemesis model (Melott \& Bambach, 2010, 2013) since the orbital period of
Nemesis is expected to increase
 by 20\%  over only 250 Myr  due to stellar and galactic tidal perturbations (Hut, 1984). The variation in period may be
further complicated due to the galactic
 radial migration  of the sun by several kpc over the last few Gyr (Kaib, et al., 2011).

The second astronomical model associates the extinction period with the vertical galactic oscillations of the solar
system. However, this model
 has been challenged on the
 basis of an inconsistency in phasing and period.  The period of galactic plane crossings ($\sim$ 33 Myr) in this model
appears to be
 inconsistent with the 27 Myr period,
 unless a significant thin disk of weakly interacting dark matter
is postulated to lie in the galactic plane (Rampino, 2015). The most recent plane crossing was $\sim$ 1 Myr ago (Gies \&
Helse, 2005) while the most recent
predicted periodic extinction event was $\sim$ 11 Myr ago (Melott \& Bamback, 2010).

The third proposed astronomical model is Planet X (Whitmire \& Matese, 1985; Matese \& Whitmire, 1986a, 1986b). In this
model the extinction period is
associated with the perihelion/aphelion precession period of a trans-Neptunian planet of mass  $\sim$ 5 M$_{\oplus}$. The
planet's perihelion and
aphelion points pass through the ecliptic plane twice during each circulation period, triggering comet showers.
 The original model
required the existence of the Kuiper belt (yet to be discovered) and a gap or edge in the belt like the now well-known
Kuiper Cliff at $\approx$ 50 AU.
The model also necessarily produces a flux of short period comets. Though in the
original model we focused on the idea that Planet X itself created the required gap, implying a mass range of 3 - 8 M$_{\oplus}$, we noted that the gap/edge might also
be primordial (Matese \& Whitmire,
1986a), as has now been observed  around some young stars. In this case the mass of Planet X could be lower, $\sim 1$ M$_
{\oplus}$, since it need not have
 cleared the gap itself. This lower limit is based solely on the comet scattering dynamics.
Other hybrid scenarios (Lykawka and Mukai, 2008)   are also possible whereby Planet X created the gap prior to major
planet migration and
subsequent evolution to  its present orbit.

 Recently it has been noted that the arguments of perihelia ($\omega$) of the two known inner Oort cloud objects with $q
\geq 75$ AU, Sedna and 2012VP$_{113}$, (as
  well as ten other trans-Neptunian objects (TNO's) with $q \geq 30$ AU and $a \geq 150$ AU)  cluster in the ecliptic
around $\omega = 0$
  (Trujillo \& Sheppard, 2014). The absence of any clustering
   around   $\omega = 180^{\circ}$ implied that the effect was not due to observational bias. These authors showed that
the clustering of the two inner Oort cloud
   objects could be explained/maintained by a
 super-Earth planet of mass 2 - 15 M$_{\oplus}$ in a near circular, small inclination orbit between 200 and 300 AU but
noted that those orbits were not
 unique. The orbit used for illustration was that of  a 5 M$_{\oplus}$ planet at 210 AU.
 The planet is needed to {\em maintain} the clustering (by causing the $\omega$'s to librate around $\omega$ = 0)
 over the age of the solar system since the known planets would precess and  thus essentially randomize
 the $\omega$'s on timescales short  compared to the age of the solar system (Trujillo \& Sheppard, 2014). The reason for
the original asymmetry
 in $\omega$ = 0  versus 180$^{\circ}$
 is left unexplained, however see de la Fuente Marcos,  de la Fuente Marcos \& Aarseth (2015). In the simulations the smaller
perihelia objects did not cluster in the
 same way but it was noted that simulations with  planets in
 higher inclination orbits should be investigated (Trujillo and Sheppard, 2014).  de la Fuente Marcos \& de la Fuente
Marcos (2014) confirmed that the clustering
 of $\omega$'s is not  due to observational bias. They also noted a tendency
 for inclinations to cluster around 20$^{\circ}$ and conjectured that there may be two such trans-Neptunian planets. The
eccentricities of these TNO's cluster
 around $\sim$ 0.8. However, unlike the  $\omega$'s, the clustering of $e$'s is likely to be at least partly a selection
effect. Thus far there have not
 been any
 other complete explanations put forward as an alternative to the $\omega =0$ clustering other than a trans-Neptunian
super-Earth planet (C. Trujillo, personal
 communication).

\section*{Analysis}

We wish to determine under what conditions, if any, the perihelion/aphelion precession of the orbit of Planet X is
sufficiently regular to be consistent with the
regularity
found in the updated extinction period. This means that the precession period should not vary {\em systematically} by more
than 2-10\%  over 500 Myr
(Melott and Bambach, 2010, 2014), which as noted above is inconsistent with the Nemesis model.
It is not obvious that
this will naturally occur since the precession period depends on the orbit's eccentricity and inclination, which might
vary significantly over 500 Myr.
On the other hand the timing of the peak maxima can fluctuate non-systematically about the 27 Myr timeline. This could
occur because of small amplitude oscillations in eccentricity or inclination or just the statistics of shower event timings.

We use the truncated secular Lagrange equations as given by Gallardo et al. (2012) as an approximation to describe the
motion of a massless Planet X, {\it i.e.}
 we neglect any back reaction on the giant planets. This approximation will thus be more accurate for the lower end of the
Planet X mass range. These equations are based on an integral truncation of the secular Hamiltonian
 and as such will not exhibit the full complexity of the dynamics.
Mean motion resonances with Neptune are also neglected in the present analysis.  With these caveats, the secular truncated
orbital equations are then:

\begin{equation}  \label{eq:xdef}
\frac{da}{dt}=0
\end{equation}

\begin{equation}  \label{eq:xdef}
\frac{de}{dt}= \frac{45ekE}{512a^{\frac{11}{2}}(1-e^2)^3} (5+7\cos2i)\sin^{2}i \sin2\omega +\dots
\end{equation}

\begin{equation} \label{eq:xdef}
\frac{di}{dt}= \frac{45ekE}{1054a^{\frac{11}{2}}(1-e^2)^4} (5+7\cos2i)\sin2i \sin2\omega +\dots
\end{equation}

\begin{equation} \label{eq:xdef}
\frac{d\Omega}{dt} = -\frac{3Ck}{4a^{\frac{7}{2}}(1-e^2)^2}\cos i +\dots
\end{equation}

\begin{equation} \label{eq:xdef}
\frac{d\omega}{dt} =\frac{3Ck}{4a^{\frac{7}{2}}(1-e^2)^2}(3+5\cos2i) +\dots
\end{equation}

In these equations $e$ is the orbit eccentricity, $i$ is the orbit plane inclination relative to the ecliptic, $\Omega$ is
the angle of the orbit's ascending
node, $\omega$ is the argument of perihelion, and $a$ is the semimajor axis. The constants $C$ and $E$ are the quadrupole
and octopole strengths equal to 0.116
and 52, respectively (Gallardo et al., 2012), and $k$ is the Gaussian constant = $\sqrt{GM_{\odot}} = 2\pi$ in units
where time is measured in years,
distances in AU, and masses in solar units. We assume that the Hamiltonian is approximately only a function of the
actions, which will be true if the Kozai harmonic, present in Eqs. (2) and (3), circulates rapidly, in which case $e$ and
$i$ are approximately constant. The precession period is then found from Eq. (5), $d\omega/dt = 2\pi/T_{\omega}$, or

\begin{equation} \label{eq:xdef}
T_{\omega} \approx \frac{16(1-e^2)^{2}a^{\frac{7}{2}}}{3C(3+5\cos2i)}
\end{equation}
 In the Planet X model the precession period = 2$\times$ 27 Myr = 54 Myr is fixed. The required semimajor axis is then
given by
\begin{equation}\label{eq:xdef}
a \approx \Big[\frac{3CT_{\omega}(3+5\cos2i)}{16(1-e^2)^2}\Big]^{\frac{2}{7}}
\end{equation}

Equation (7) predicts a unique relation between $i,e,$ and $a$, though for a given $T_{\omega}$ period the required
semimajor axis is fairly insensitive to
$e$ and $i$.  If showers at perihelion dominate then the eccentricity is $\leq$ 0.5, while for aphelion showers  there is no
limitation on the
eccentricity of the orbit other than that the perihelion point not get too close to Neptune's orbit (Matese and Whitmire,
1986a). The required inclination needs
to be $\geq 25^{\circ}$ in order that comets can be directly scattered into the spheres of influence of Jupiter and
Saturn, as required by the duration
of the shower in the inner solar system. Though the 27 Myr extinction period is highly regular, the duration or width of
the shower is less well constrained.

Gallardo, et al. (2012) investigated  Kozai dynamics as applied to
trans-Neptunian bodies, considering both
non-resonant and mean motion resonances. We focus on their Fig. 2, panels (e) - (h), which give the non-resonant curves of
$q$ and $i$ versus $\omega$ for
$a$ = 100 AU and various values of the constant $H$ = $\sqrt{1 - e^2}\cos i$ from 0.1 to 0.7. Because of the symmetries,
$\omega$ and $i$ are shown only to 180$^{\circ}$ and 90$^{\circ}$, respectively.
Inspection of these figures show that for a given
$q \geq$ 50 AU (Kuiper Cliff), corresponding to approximately constant/equilibrium inclinations in the range of 36$^
{\circ} - 83^{\circ}$, neither
the perhelia, eccentricites or inclinations vary by more than a few percent as $\omega$ takes on all values in its circulation.
Modest increases of H $\geq$ 0.7 and/or $a \geq 100$ AU would allow
values of $i$ down to 25$^{\circ}$, which is the dynamical
scattering limit of the Planet X model.
For $a$ = 100 AU, values of $q \leq$ 50 AU would be incompatible with the cold classical Kuiper belt which has an inclination
dispersion of $\approx 5^{\circ}$ (Brown 2001).

\section*{Discussion}

Inserting $e$ = 0.5 and $i$ = 25$^{\circ}(35^{\circ})$ into Eq. (7) gives $a$ = 108(100) AU, $q$ = 54(50) AU and $Q$ = 162(150) AU.  These
aphelia are less than the 200 - 300 AU near
circular, low inclination orbits
assumed by Trujillo and Sheppard in most of their simulations but, as noted above, they acknowledged that their masses and
orbits are not unique in maintaining
the clustering
about $\omega$ = 0. In their simulations the planet masses included the super-Earth range of 2 - 15 M$_{\oplus}$ at 200 -
300 AU. In their illustration they used
a 5 M$_{\oplus}$ planet at 210 AU. They assumed that with a very low albedo such planets could have remained undetected in
prior full sky surveys, presumably
even near the ecliptic.
In the Planet X model the conservative mass limit was given as 1 - 15 M$_{\oplus}$ (Matese and Whitmire, 1986a), though
assuming Planet X itself created the gap
the range given  was 3 - 8 M$_{\oplus}$. However, as noted above, and in our paper, if the Kuiper Cliff is primordial a mass as low as 1 M$_{\oplus}$
is possible, based solely on the scattering dynamics.
A low albedo planet with a mass near the lower end of the range with inclination $\geq 25^{\circ}$  and currently near
apehelion might likewise have escaped detection  in past full sky surveys.

Tombaugh's full sky survey that discovered Pluto is still relevant beyond the ecliptic and certain other limited  deep
field studied regions.
His full sky search was sensitive to $\approx$ 15th
magnitude. The visual magnitude of Pluto at the time of discovery was $\approx$ 14.5. The reflected flux of Planet X at
Earth relative to that of Pluto is
$\propto (A_{X}/A_{P})(M_{X}/M_{P})^{\frac{2}{3}}(r_{P}/r_{X})^{4} $, where $A$ is albedo, $r$ is the heliocentric
distances, equal densities
are assumed, and the Earth-sun distance is neglected.
Inserting Pluto's mass = 2.2$\times 10^{-3}$ M$_{\oplus}$, $M_{X} = 2$ M$_{\oplus}$,  $r_P = 35$ AU and $r_X$ = $Q$ =150 AU
gives a
relative reflected flux =
0.278. This is 1.4 magnitudes dimmer than Pluto, or $\approx$ 16th magnitude, assuming equal albedos. Pluto's average
albedo is 0.58.
If Planet X has a low albedo of say 0.05 then the
relative reflected flux would be further reduced to 0.024, making it about 4 magnitudes dimmer than Pluto, or magnitude
18.5. A planet of visual
magnitude 16 - 18.5 currently located well out of the ecliptic could have been missed in Tombaugh's full sky survey and
other subsequent limited-field optical
 surveys. Luhman (2014) does not give WISE full sky mid-IR limits for rock/ice trans-Neptunian planets.

Lykawka and Mukai (2007) argued that a trans-Neptunian planet of mass up to 1 M$_{\oplus}$ can explain the architecture of
the Kuiper belt.
Some features like the Kuiper Cliff at $\sim$ 50 AU were explained by the planet prior to outer-planet-migration. The
planet subsequently evolved to
its present orbit, $a$ $\geq$ 100 AU and $i = 20^{\circ} - 40^{\circ}$. In this analysis the planet is predicted to have
$q \geq 70 - 80$ AU and they
used this result to rule out several trans-Neptunian planet predictions, including the Planet X model considered here.
However, even if this constraint on $q$ is correct,
the Planet X model can still accommodate it.  For example setting $i = 25^{\circ}$ and $e$ = 0.2(0.1) in Eq. (7) gives $a$
= 93.5(91.6) AU and $q$ = 74.8(82.4) AU.

Bromley \& Kenyon (2014) modeled the fate of  planets scattered by more massive planets as the scattered planet interacted
with a proto-planetary disk. Various
planet masses and disk models were studied. Scattered planets with masses $\geq$ 5 M$_{\oplus}$ tended to be circularized
by the disk while  planets of mass
$\leq$ 2 M$_{\oplus}$ maintained their original scattered eccentricity and inclination. These authors note that a remote
planet in the solar system
at $\sim$ 100 AU is plausible and might explain the $\omega$ clustering.

In response to the papers by Trujillo \& Sheppard  (2014) and de la Fuente Marcos \& de la Fuente Marcos (2014), Iorio
(2014) argued that a super-Earth planet
at 250 AU would induce precessions in the known planets' orbits that are incompatible with observations. However, the
proper approach to this problem requires
integrations with and without the hypothesized planet (Hogg et al., 1991;
C. de la Fuente Marcos, personal communication;
S. Tremaine, personal communication). It is problematic to base conclusions on residuals from a best fit model with many
parameters that don't include the
unknown planet (Hogg et al., 1991). That is, Iorio's approach doesn't account for the equivalence between the effects of a
new planet and other adjustable
parameters in
the fit.  Further, the limits found by Iorio are an order of magnitude more restrictive than earlier constraints (Tremaine
1990; Hogg et al., 1991;
Zakamska and Tremaine, 2005)
even though the ephemerides are not that much better (S. Tremaine, personal communication).

In addition to the highly regular 27 Myr period an even stronger 62 Myr period has also been identified in the fossil record (Rohde \& Muller, 2005; Melott \& Bambach,
2007, 2013, 2014). One astronomical idea to explain this longer period is an increase in the cosmic ray flux as the solar
system oscillates to high northern
galactic latitudes (Lieberman and Melott, 2012).
Another possibility is  that a similar Planet X model-like mechanism is at work involving a second planet. However, the
dynamics of such a system cannot be addressed using the limited approximations of the present analysis since planet-planet interactions may dominate the precession frequencies. Alternatively, a more complete dynamical analysis including resonances with the giant planets
might conceivably lead to a single planet
explaining both periods.

\section*{Conclusion}

With modest restrictions on inclination and eccentricity the Planet X model, as originally presented, is shown to be
consistent with the regularity of the fossil
extinction period, in contrast to the Nemesis model.  Recent observational and theoretical studies suggest that a trans-
Neptunian planet is more plausible
today than when the model was first proposed. Although Trujillo \& Sheppard (2014) acknowledged that their simulated
orbits and masses
were not unique, further numerical studies are needed to investigate the phase space of  orbital parameters and
masses capable of maintaining the $\omega$ clustering of detached TNO's around $\omega$ = 0. Of special interest is a
planet of a few Earth-masses, semimajor
axis $\sim$ 100 AU, in an orbit with moderate eccentricity and inclination $\geq 25^{\circ}$.

\section*{Acknowledgments}

I thank Carlos de la Fuente Marcos and Scott Tremaine for comments on dynamical limits applied to trans-Neptunian planets
and John Matese and Chad Trujillo for other helpful comments. I also thank an anonymous  referee for a critical review and suggestions that significantly improved the manuscript.

\end{document}